\documentstyle[12pt]{article}

\newcommand{\ds}{\displaystyle}
\newcommand{\la}[1]{\label{#1}}
\newcommand{\re}[1]{\ (\ref{#1})}
\newcommand{\nn}{\nonumber}
\newcommand{\ed}{\end{document}}
\newcommand{\be}{\begin{equation}}

\newcommand{\ee}{\end{equation}}   \newcommand{\ba}{\begin{eqnarray}}

\newcommand{\baa}{\begin{array}{lll}}   \newcommand{\eaa}{\end{array}}
\newcommand{\ea}{\end{eqnarray}}
\newcommand{\baz}{\begin{eqnarray*}}
\newcommand{\eaz}{\end{eqnarray*}}
\newcommand{\bb}{\begin{thebibliography}}
\newcommand{\eb}{\end{thebibliography}}

\newcommand{\bi}[1]{\bibitem{#1}}

\begin{document}
\phantom{.}
{\hfill                                     JINR, E2-97-54}

{\hfill                                     hep-ph/9702417}
\vspace {3cm}
\begin{center}
\large{VIRTUALITIES OF QUARKS AND GLUONS IN QCD VACUUM\\
AND NONLOCAL CONDENSATES
\\ WITHIN SINGLE INSTANTON APPROXIMATION.}\\[0.5cm]

{A.~E.~Dorokhov$^{\star}$
\footnote{E - mail: dorokhov@thsun1.jinr.dubna.su},
S.V. Esaibegyan$^{\star,\ast}$,
S.V. Mikhailov$^{\star}$
\footnote{E - mail: mikhs@thsun1.jinr.dubna.su}}\\[0.5cm]

$^{\star}${\it
Joint Institute for Nuclear Research,\\
Bogoliubov Theoretical Laboratory,\\
141980, Moscow Region, Dubna, Russia}\\[0.5cm]

$^{\ast}${\it Yerevan Physics Institute, 375036, Yerevan, Armenia}\\[0.5cm]

 Abstract\\[0.3cm]
\end{center}
We calculate the lowest-dimensional nonlocal quark and gluon
condensates within the single instanton approximation of the instanton
liquid model.  As a result,
we determine the values of average virtualities of quarks
$\lambda_q^2$ and gluons $\lambda_g^2$ in the QCD vacuum and obtain
parameterless predictions for the ratio $\lambda_g^2/\lambda_q^2 =
12/5$, and for some ratios of different vacuum condensates of higher
dimensions.  The nonlocal properties of quark and gluon condensates
are analyzed, and insufficiency of the single instanton approximation
is discussed.

\section{Introduction}

The nonperturbative vacuum of QCD is densely populated by long - wave
fluctuations of gluon and quark fields. The order parameters of this
complicated state are characterized by the vacuum matrix elements of
various singlet combinations of quark and gluon fields, condensates:
$<:\bar qq:>$,~$<~:~G^a_{\mu\nu}G^a_{\mu\nu}:>$, $<:\bar
q(\sigma_{\mu\nu}G_{\mu\nu}^a\frac{\lambda^a}{2})q:>$, {\it etc}.
The nonzero quark condensate $<:\bar qq:>$ is responsible for the
spontaneous breakdown of chiral symmetry, and its value was estimated
a long time ago within the current algebra approach. The importance
of the QCD vacuum properties for hadron phenomenology have been
established by Shifman, Vainshtein, Zakharov \cite{SVZ79}. They used
the operator product expansion (OPE) to relate the behavior of hadron
current correlation functions at short distances to a small set of
condensates. The values of low - dimensional condensates were
obtained phenomenologically from the QCD sum rule (QCD SR) analysis
of various hadron channels.

Values of higher - dimensional condensates are known with less
accuracy since usually in the range of applicability of QCD SR the
static hadron properties:  lepton widths, masses, {\it etc,} are less
sensitive to respective corrections. The whole series of power
corrections characterizes the nonlocal structure of vacuum
condensates.

Nonlocality  of the quark condensate is  characterized by the
parameter \cite{MihRad92}
\be
\lambda^2_q = \frac{<:\bar q D^2q:>}{<:\bar q q:>},
\la{Lq}\ee
where $D_\mu = \partial_\mu - ig A_\mu^a \lambda^a/2$ is a covariant
derivative. This quantity is treated as average virtuality of quarks
in the QCD vacuum and characterizes the space width of quark
distribution.  By the equation of motion in the chiral limit the
parameter $\lambda^2_q$ is also related to the mixed quark - gluon
condensate
\be
m_0^2 = \frac{\ds <:\bar
q(ig \sigma_{\mu\nu}G_{\mu\nu}^a\frac{\lambda^a}{2})q:>}{\ds <:\bar q q:>},
~~~~~~~\lambda^2_q = \frac{m_0^2}{2}.
\la{M0}\ee
This quantity has been estimated by QCD SR for baryons to $m_0^2 =
0.8 \pm 0.2\ GeV^2$ \cite{BI82}, and the lattice QCD (LQCD)
calculations yield $m_0^2 = 1.1 \pm 0.1\ GeV^2$ \cite{KSch87}. Within
the instanton model the mixed condensate has first been obtained in
the single instanton approximation in \cite{Sh82}
\footnote{ The result given in
\cite{Sh82}  differs from a correct one by factor $1/2$. } .
Recently, similar
calculations  has been performed in a more advanced instanton vacuum
model \cite{PolW96} with the result
$m_0^2 \approx \frac{\ds 4}{\ds \rho_c^2}$, where $\rho_c$ is the
characteristic size of the instanton fluctuation in the QCD vacuum.
Below, we reproduce this result in another way. As for the
nonperturbative properties of gluons in the QCD vacuum, new precise
LQCD measurement of the gauge - invariant bilocal correlator of the
gluon field strengths has become available down to a distance of
$0.1\ $ fm\ \cite{DiGicor96}.

As it has been proposed in \cite{MihRad92}, the nonlocal properties
of vacuum condensates are of principal importance in the study of the
distribution functions of quarks and gluons in hadrons.  There, it
has been shown that this problem can be correctly considered only if
a certain nonlocal form of the vacuum condensates is suggested.
Physically, it means that vacuum quarks and gluons can flow through
the vacuum with nonzero momentum.  To construct the simplest ansatzes
for the shape of the nonlocal condensates, in \cite{MihRad92,Mih93}
some general properties of these functions and the restricted
information about their first derivatives have been used.

On the other hand, in QCD there is an instanton \cite{R1,R4}, a well
- known nontrivial nonlocal vacuum solution of the classical
Euclidean QCD field equations with the finite action and size $\rho$.
The importance of instantons for QCD is that it is believed that an
interacting instanton ensemble provides a realistic microscopic
picture of the QCD vacuum in the form of ``instanton liquid''
\cite{Sh82,Sh88,DP84} (see, {\it e.g.}, a recent review
\cite{Shuryak96}). It has been argued on phenomenological grounds
that the distribution of instantons over sizes is peacked at a
typical value $\rho_c\approx 1.7\ GeV^{-1}$ and the ``liquid'' is
dilute in the sense that the mean separation between instantons is
much larger than the average instanton size. Moreover, the quark
Green functions are dominated by zero energy modes localized around
the instanton. The effects of condensate nonlocality within the
instanton liquid model have implicitly been used in QCD SR for the
pion \cite{Sh83} and nucleon \cite{DKsr90}, where they appear as
exponential corrections to the sum rules along with power corrections
typical of the local OPE approach.

In this paper, we start a systematic discussion of nonlocal
condensates within the instanton model of the QCD vacuum.  As a first
step, we calculate average virtualities of quarks and gluons in the
QCD vacuum in the single instanton (SI) approximation. Next, we
attempt to obtain the correlation functions $f(\nu)$ which describe
distributions over virtuality $\nu$ of quarks and gluons in the
nonperturbative vacuum. The approximation used works well for large
virtualities, but fails in the description of physically argued
distributions at small virtualities (or long distances). The reason
is that in order to have a realistic model of vacuum distributions,
the important effects of long - wave vacuum configurations  have to
be included~\cite{DEMM296}.

The paper is organized as follows. In the second section, the general
properties of nonlocal condensates are briefly discussed. The quark
and gluon average virtualities $\lambda^2$ are estimated  within the
single instanton approximation in the third section. To guarantee the
gauge invariance, we have introduced the Schwinger $\hat E -$
exponent as an operator element of the nonlocal vacuum averages. In
the fourths section, we analyze the space coordinate behavior of
nonlocal condensates. The main asymptotics of the correlation
functions $f(\nu)$ at large virtualities $\nu$ are derived. We also
demonstrate insufficiency of the SI approximation to obtain the
realistic behavior at large distances. There, we point out the
physical reason for the failure of the approach used in the large
distance region and suggest a way to solve this problem.

\section{The quark and gluon distribution functions in the QCD vacuum}

To begin, we outline some basic elements of the approach with the
nonlocal vacuum condensates.  The simplest bilocal scalar condensate
$M(x)$ or, in other words, the nonperturbative part of the gauge -
invariant quark propagator has the form (in the below definitions we
shall follow works \cite{MihRad92,Mih93})
\begin{equation}
M(x)\equiv<:\bar {q}(0)\hat{E}(0,x)q(x):>\equiv<:\bar {q}(0)q(0):> Q(x^2).
\label{E18}
\end{equation}
Here, $\hat{E}(x,y)=P\exp \left( i \int^y_x A_\mu(z)dz^{\mu} \right)$
is the path-ordered Schwinger phase factor (the integration is
performed along the straight line) required for gauge invariance and
$\ds A_{\mu}(z) = gA_{\mu}^a(z) \frac{\lambda^a}{2}$.  In the same
manner, we will consider the correlator $D^{\mu\nu,\rho\sigma}(x)$ of
gluonic strengths $\ds G_{\mu\nu}(x) = gG_{\mu\nu}^a(x)
\frac{\lambda^a}{2}$
\begin{equation}
D^{\mu\nu,\rho\sigma}(x-y)\equiv
<:Tr G^{\mu\nu}(x)\hat{E}(x,y)G^{\rho\sigma}(y)\hat{E}(y,x):>.
\label{Mg}\ee
The correlator may be parameterized in the form consistent with
general requirements of gauge and Lorenz symmetries as
\cite{DoSi88,Mih93,Groz}:
\ba
D^{\mu\nu,\rho\sigma}(x) & \equiv & \frac{1}{24} <:g^2 G^2:> \{
(g_{\mu\rho} g_{\nu\sigma} - g_{\mu\sigma} g_{\nu\rho})
[D(x^2) + D_1(x^2)]  +
\la{GNL}\\
& + & (x_\mu x_\rho g_{\nu\sigma} - x_\mu x_\sigma g_{\nu\rho} +
x_\nu x_\sigma g_{\mu\rho}  - x_\nu x_\rho g_{\mu\sigma} )
\frac{\partial D_1(x^2)}{\partial x^2}           \},
\nn\ea
where $<:G^2:>=<:G^a_{\mu\nu}(0)G^a_{\mu\nu}(0):>$ is a gluon
condensate, and $Q(x^2)$, $D(x^2)$ and $D_1(x^2)$ are invariant
functions that characterize nonlocal properties of condensates.

The vacuum expectation values (VEV) like $<:\bar {q} q:>$, $<:g^2
G^2:>$, ~$<:\bar {q} D^2 q:>,\ \ldots $ appear as expansion
coefficients of the correlators $M(x)$ and $D^{\mu\nu,\rho\sigma}(x)$
in a Taylor series in the variable $x^2/4$.  The coordinate
dependence of the scalar condensates $Q(x^2)$ and $D(x^2)$,
normalized at zero by $Q(0) = 1$ and $D(0) + D_1(0) = 1 $, can
conventionally be parameterized similarly to the well - known
$\alpha$-representation for the propagator \footnote{One has to
remember that in this work we make use of the Euclidean space and
$x^2 < 0$.}
\ba
\ds Q(x^2)&=&\int_{0}^{\infty} \exp\left(\frac{x^2}{4\alpha} \right)
f_q\left(\frac{1}{\alpha}\right) {d\alpha \over \alpha^2}, \label{fq}\\ \ds
D(x^2)+D_1(x^2)&=&\int_{0}^{\infty} \exp\left(\frac{x^2}{4\alpha}
\right) f_g\left(\frac{1}{\alpha}\right) {d\alpha \over \alpha^2}.
\label{fg} \ea

The properties and the role of the correlation functions $f(\nu)$
have been discussed in detail in \cite{MihRad92,Mih93}. The explicit
form of $f(\nu)$ completely fixes the coordinate dependence of the
condensates and can be determined in the future QCD vacuum theory.
Evidently, $f(\nu) \sim \delta(\nu),\ \delta^{\{1\}}(\nu),$\ ...,
would correspond to the standard VEVs $<:\bar qq:>$, $m^2_0$,\ ...,
while the behavior $f(\nu) \sim const$ would simulate free
propagation. We expect that the realistic $f(\nu)$ occurs somewhere
in between these two extremes. Thus, it is a continuous function
concentrated around a certain finite value $\lambda^2$ and rapidly
decaying to zero as $\nu$ goes to $0$ or $\infty$.

The correlation function $f_q(\nu)$ describes the virtuality
distribution of quarks in the nonperturbative vacuum \cite{MihRad92}.
Its $n$-moment is proportional to the VEV of the local operator with
the covariant derivative squared $D^2$ to the $nth$ power
\begin{equation}
\int_{0}^{\infty}\nu^{n}f_q(\nu)d\nu =\frac{1}{\Gamma(n+2)}
\frac{<:\bar {q}(D^2)^{n}q:>}{<:\bar {q}q:>}.
\label{E20}
\end{equation}
It is natural to suggest that VEVs in the {\it r.h.s.} of \re{E20}
should exist for any $n$.  It means that the decrease of $f(\nu)$ for
large arguments has to be faster than any inverse power of $\nu$,
{\it e.g.}, like some exponential
\be
f_q(\nu) \sim \exp(-const \cdot \nu)\ \ \  \mbox{as}\ \ \ \nu \to
 \infty.  \la{flarge}\ee

The two lowest moments give the normalization conditions and the
average vacuum virtualities of quarks $\lambda^2_q$ and gluons
$\lambda^2_g$
\ba
&&\int_{0}^{\infty} f_q(\nu)
d\nu=1,~\int_{0}^{\infty} \nu f_q(\nu) d\nu= \frac{1}{2}\frac{<:\bar {q}
D^2 q:>}{<:\bar {q}q:>} \equiv \frac{\lambda^{2}_{q}}{2},
~~(\lambda^{2}_{q}\simeq 0.4 \ GeV^2,
 \mbox{QCD SR \cite{BI82})},
\label{E21} \\
&& \int_{0}^{\infty}
f_g(\nu) d\nu=1,~\int_{0}^{\infty} \nu f_g(\nu) d\nu=
  \frac{1}{2}\frac{<:G^a_{\mu\nu} D^2 G^a_{\mu\nu}:>}{<:G^2:>} \equiv
  \frac{\lambda^{2}_{g}}{2}.
\label{Lg}
\ea

Note that the quark correlator \re{E18} has a direct physical
interpretation in the heavy quark effective theory (HQET) of heavy -
light mesons as it describes the propagation of a light quark in the
color field of an infinitely heavy quark \cite{Sh82,Shuryak96}. This
behavior has been analyzed in detail in \cite{Rad91}. There, it was
demonstrated  that for large distances $|x|$ the correlator is
dominated by the contribution of the lowest state of a heavy - light
meson with energy $\Lambda_q$:  ~$Q(x^2) \sim \exp\left(- \Lambda_q
|x| \right)$. This law provides the behavior of $f(\nu)$ at small
$\nu$
\be
f_q(\nu) \sim \exp(-\Lambda_q^2 / \nu) \ \ \ \mbox{as}\ \ \ \nu \to 0.
\la{fsmall}\ee
In the case of gluon correlator (\ref{GNL}) the correlation length
$L_g$ has recently been estimated in the LQCD calculations
\cite{DiGicor96}.  The quantity $\Lambda_g=1/L_g$ plays a similar
role as $\Lambda_q$ for the quark distribution, {\it i.e.}~$\ds D(x^2) \sim
\exp\left(-\Lambda_g |x| \right)$ for large $|x|$.  It is formed at typical
distances of an order of $0.5 \ {\rm fm}$ and describes long range vacuum
fluctuations of gluon field.

In works \cite{BM95,BM96}, the arguments in favor of a definite
continuous dependence of $f(\nu)$ have been analyzed and different
ansatzes for these functions  were suggested which are consistent
with the requirements \re{flarge}, \re{fsmall}. In particular, one
ansatz has been constructed by the simplest combination of both these
asymptotics
\be
 f_q(\nu) \sim \exp\left(-\frac{\Lambda_q^2}{\nu}-\sigma_q^2 \cdot \nu\right)
\label{E22}
\end{equation}
with the parameters $\Lambda_q \simeq 0.45~\mbox{GeV}$ and
$\sigma_q^2 \simeq 10\ \mbox{GeV}^{-2}$.  This ansatz has been
successfully applied in QCD SR for a pion and its radial excitations
~\cite{BM96}, and the main features of the pion have been described:
the mass spectrum of pion radial excitations $\pi'$ and $\pi''$ which
is in agreement with the experiment and the shapes of the wave
functions of $\pi$  and $\pi'$ which have been confirmed by an
independent analysis in ~\cite{MihRad92,BM95}.  Thus, we will regard
the form (\ref{E22}) as following directly from the pion
phenomenology.  Below, we will make some conclusions about the form
of the correlation function $f_q(\nu)$ using concrete solutions for
the instanton field and quark zero mode around it.

\section{Vacuum average virtualities in the single instanton approximation}

Let us consider an instanton solution of the classical Yang - Mills
equations in the Euclidean space \cite{R1}. It is well known that in
the vicinity of the instanton the quark amplitudes are dominated by
the localized mode with zero energy \cite{R4}. We will consider the
expressions for the instanton field and quark zero mode in the axial
gauge $A_\mu(z) n^\mu =0$ since in this gauge with the vector $n_\mu
= x_\mu - y_\mu$ the Schwinger factor $\hat{E}(x,y)=1$.  The
expressions in the axial gauge for the instanton $(+)$
(anti-instanton $(-)$) field

\be A_{\mu(ax)}(x) = R(x)A_{\mu(reg)}R(x)^+ +
iR(x)\partial_\mu R(x)^+, \ \ \ \ \  G_{\mu\nu(ax)}(x)  =
R(x) G_{\mu\nu(reg)}R(x)^+, \la{Aax}\ee
and the quark zero mode
\ba
\Psi^\pm_{ax}(x) = R_\pm(x) \Psi_{reg}^\pm(x),
\la{qax}
\ea
where
\ba
R_\pm(x) =   \exp\left[\pm i (\vec{x} \vec{\tau}) \alpha(x) \right],\ \ \ \
\alpha(x)=\frac{|\vec x|}{\sqrt{x^2+\rho^2}}
 \arctan\frac{x_{4}}{\sqrt{x^2+\rho^2}}
\nn\ea
have been introduced in \cite{MHax}. In \re{Aax} and \re{qax} the
expressions for the instanton and quark fields in the regular gauge
are given by
\ba
 A_{\mu,reg}^{\pm a}(x) & = &\eta^{\mp a}_{\mu\nu}
 \frac{2x_{\nu}}{x^2+\rho^2},
\ \ \ \ G^{\pm a}_{\mu\nu,reg}(x) =
- \eta^{\mp a}_{\mu\nu}\frac{4 \rho^2}{(x^2+\rho^2)^2},
\la{Areg}\\
\Psi_{reg}^\pm(x) & = & \varphi_{reg}(x) \xi^\pm ,\quad\quad
 \varphi_{reg}(x) =  \frac{\rho}{\pi(x^2+\rho^2)^{3/2}}.
\la{qreg}\ea

In \re{Aax} - \re{qreg},  $x = (x_4,\ \vec x)$ is a relative
coordinate with respect to the position  of the instanton center $z$.
 The solutions \re{Aax} and \re{qax} are given within the $SU(2)$
subgroup of the $SU_c(3)$ theory ($\tau_a$ are the corresponding
generators normalized according to $Tr(\tau_a \tau_b) = \frac{1}{2}
\delta^{ab}$) and the following notation is introduced:  $\ds
\eta^{a\pm}_{\mu\nu} = \epsilon_{4a\mu\nu} \pm
\frac{1}{2}\epsilon_{abc}\epsilon_{bc\mu\nu}$ are the t'Hooft
symbols, $\ds \xi^{\pm}\bar\xi^{\pm}=\frac{1}{8}\gamma_{\mu}\gamma_{\nu}
\frac{1\pm{\gamma_5}}{2}U \tau^{\mp}_{\mu}\tau^{\pm}_{\nu} U^{+}$
with $\tau^\pm = (\pm i,\vec \tau)$, and $U$ is the matrix of color
space rotations.

In the SI background in the zero  mode approximation the bilocal
quark  and gluon condensates acquire the form
\ba
\ds M_q(x)& = & <:\bar{q}_{(ax)}(0)q_{(ax)}(x):>  = \nonumber \\
\ds & = & -\sum_{\pm}n_{c}^\pm\int d^4{z}\int d\Omega
\frac{Tr[\Psi^{\pm}_{ax}(x-z) \bar
{\Psi}^{\pm}_{ax}(-z)]}
{m^{*}_q},
\label{E23} \\
D^{\mu\nu,\rho\sigma}(x) &=&
<:G_{(ax)\mu\nu}^a(0)G_{(ax)\rho\sigma}^a(x):> =
\nn \\
\ds &=&\frac{1}{12} (g_{\mu\rho} g_{\nu\sigma} - g_{\mu\sigma} g_{\nu\rho})
 \sum_{\pm}n_{c}^\pm\int d^4{z}\int d\Omega
G^{\pm a}_{\delta\delta'(ax)}(x-z)G^{\pm a}_{\delta\delta'(ax)}(-z).
\label{Mgax}
\ea
Here, $n_c^\pm$ is the effective instanton / anti - instanton
density.  The collective coordinate $z$ of the instanton  center and
its color space orientation are integrated over. In the SI
approximation the term in \re{GNL} with the second Lorenz structure
does not appear. This fact is due to the specific topological
structure of the instanton solution. Both the Lorenz structures will
appear in the r.h.s. of (\ref{Mgax}) if one takes into account the
long-wave background fields \cite{DEMM296}.

The averaging over the instanton orientations in the color space is
carried out by using the relation $\ds \int d\Omega
U^{a}_{b}U^{+c}_{d}  =  \frac{1}{N_{c}}\delta^{a}_{d}
\delta^{c}_{b}$, where $N_c$ is the number of colors.  Using the
definitions \re{E18} - \re{GNL} and \re{E23}, \re{Mgax} we obtain
\begin{eqnarray}
\ds Q_{ax}(x^2) & = &
\frac{8\rho^2}{\pi} \int_0^\infty\ dr r^2 \int_{-\infty}^\infty\ dt
\frac{\cos[\frac{r}{R}(\arctan( \frac{t+|x|}{R}) - \arctan
(\frac{t}{R}))]} {[R^2+ t^2]^{3/2}[R^2+ (t+|x|)^2]^{3/2}}, \label{E24}\\
\ds D_{ax}(x^2) & \equiv & D(x^2)+D_1(x^2)  =  \label{Dgax}\\
\ds & = & \frac{24\rho^4}{\pi}
\int_0^\infty\ dr r^2 \int_{-\infty}^\infty\ dt
\frac{1 - \frac{4}{3} \sin^2[\frac{r}{R}(\arctan{(\frac{t+|x|}{R})}
- \arctan{(\frac{t}{R})})] }
{[R^2+ t^2]^{2}[R^2+ (t+|x|)^2]^{2}},
\nn\end{eqnarray}
where $R^2 = \rho^2 + r^2, ~r=|\vec{z}|, ~t=z_4$.
In the derivation of these equations we have used a reference frame
where the instanton sits at the origin and $x^\mu$ is parallel to one
of the coordinate axes, say $\mu = 4$, serving as a ``time" direction
$({\it i.e.},\vec x = 0,\ x_4 = |x|)$.  Expression \re{E24}
corresponds to that derived in \cite{Sh82,MHax} and expression
\re{Dgax} was derived in \cite{BP82}\footnote{We are grateful
to A. Radyushkin who communicated us this reference.}.

In the derivation of \re{E24} and \re{Dgax} the following relations
between the quark and gluon condensates, on the one hand, and the
effective density $n_{c}=n^+ + n^-$, $(n^+ = n^-)$ and the effective
quark mass $m_q^*$, on the other hand, have been used
\begin{equation}
<:\bar {q}(0)q(0):> =-\frac{n_{c}}{m^{*}_q},\
\ \ \ \ <:g^2G^2:> = 32 \pi^2 n_{c}.
\label{E12} \end{equation}
These relations are valid in the mean field approximation of the
instanton liquid model \cite{Shuryak96} and provide the normalization
conditions in \re{E23} and \re{Mgax}.  Let us emphasize two features
of expressions (\ref{E24}) and (\ref{Dgax}).

First, it is important that the factors $\cos(...)$ or $\sin^2(...)$
in the numerator of integrands reflect the presence of the $\hat{E}$
factor in the definition of the bilocal condensates.

Second, the correlators $Q(x^2) = Q_{ax}(x^2)$ and $D_{ax}(x^2)$ are
gauge - invariant objects by construction. Therefore, the same
expressions for the correlators can be derived using any other gauge.
But the axial gauge used seems to be the most adequate in this case.

From  \re{E24} and \re{Dgax} one may derive  the average virtualities
of vacuum quarks and gluons in the SI approximation which
characterize the behavior of nonperturbative propagators at short
distances in the instanton field
\begin{equation}
\lambda^2_{q}=-8\frac{dQ_{ax}(x)}{dx^2}=2\frac{1}{\rho_c^2},\ \ \
\lambda^2_{g}=-8\frac{dD_{ax}(x)}{dx^2}=\frac{24}{5}\frac{1}{\rho_c^2},\ \
\ \lambda^2_{g} = \frac{12}{5} \lambda^2_{q}.
\label{E25} \end{equation}
In expressions \re{E25} for $\lambda^2$, factor $8$ arises from the
expansion  of correlators in the variable $x^2/4$ and also due to the
definition of $\lambda_{q(g)}^2$, \re{E21} and \re{Lg}.
The result for $\lambda^2_q$ in \re{E25} agrees
with the value for the mixed condensate derived in \cite{PolW96}
if the relation \re{M0} is used.

We see that our result coincide numerically  with that derived from
the QCD SR, \re{E21}, if the effective size of the instanton is
approximately chosen as $\rho_c \approx 2$ GeV$^{-1}$
\be
\lambda^2_{q} \approx 0.5\ \mbox{GeV}^2,\ \ \ \ \
\lambda^2_{g} \approx 1.2\ \mbox{GeV}^2.
\la{LmbTheor}\ee
This value is quite close to the commonly accepted typical instanton
radii 1.7 GeV$^{-1}$ chosen to reproduce the phenomenological
properties of the instanton vacuum (see review \cite{Shuryak96}).
The recent analysis of the instanton vacuum parameters given in
\cite{PDW96} leads  to the  "window" for the $\rho_c$ value - $\rho_c
= 1.7 - 2$ GeV$^{-1}$. It is
interesting to note that gluons are distributed more compact than
quarks in the QCD vacuum as it follows from \re{E25}. To demonstrate
this it is instructive to compare the short - distance correlation
lenghts for quark $\ds l_q = \frac{1}{\lambda_q} \approx 0.28$ fm and
gluon $\ds l_g = \frac{1}{\lambda_g} \approx 0.18$ fm distributions
in the QCD vacuum ($\rho_c \approx 2$ GeV$^{-1}$).

We ignore the effects of radiative
corrections to the condensates connected with a possible change of
normalization point $\mu$ where the condensates are defined. These
effects as well as the effects due to non-zero modes contributions
are not very important.  Thus, the SI approximation works fairly well
in describing virtuality of vacuum quarks (gluons) and nonlocal
properties of condensates at short distances.  In the next section,
we are going to study the shape of nonlocal condensates in more
detail.

The relation of the quantity $\lambda^2_g$ to the
combination of VEVs of dimension six has been obtained in
~\cite{NRad83} (see also \cite{Groz}):
\be
\frac{\lambda^2_g}{2} =  \frac{<:g^3 fG^3:>}{<:g^2 G^2:>}-
\frac{<:g^4 J^2:>}{<:g^2 G^2:>},
\label{LgVA}\ee
where $<:g^3 fG^3:>= <:g^3 f^{abc}G^a_{\mu \nu}G^b_{\nu
\rho}G^c_{\rho \mu}:>$, $J^2 =  J^a_{\mu}J^a_{\mu}$ and $J^a_{\mu}=
\bar{q}(x)\frac{\lambda^a}{2} \gamma_{\mu}q(x)$.  This formula is
analogous to \re{M0} for quarks and relates short distance
characteristic \re{E25} of nonlocal gluon condensate $D(x)$ to the
standard VEVs of higher dimensions.  The estimation $\ds <:g^3 fG^3:>
\approx \frac{12}{5\rho_c^2}<:g^2 G^2:>$ following  from \re{E25} and
\re{LgVA} (without the second numerically small term) coincides with
that obtained in \cite{SVZ79,Sh82} in a different way. The latter
relation in (\ref{E25}) and the expressions for $\lambda^2_{q}$,
(\ref{Lq}), and $\lambda^2_{g}$, (\ref{LgVA}), allow us to obtain a
 new parameterless relation
\be
\frac{<:g^3 fG^3:>}{<:g^2 G^2:>} =
\frac{3}{5}\frac{<:\bar  q(ig \sigma_{\mu\nu}G_{\mu\nu}) q:>}{<:\bar qq:>}
+ \frac{<:g^4J^2:>}{<:g^2 G^2:>}, ~~(\mbox{SI approximation})
\la{dim6} \ee
and then to estimate a poorly known value of $<:g^3 fG^3:>$ $$
\frac{<:g^3 fG^3:>}{<:g^2 G^2:>} \approx  (0.45 \pm 0.12)~GeV^2. $$
To obtain this value, we have used the approximation $<:g^4J^2:>
\approx - \frac{4}{3}g^2<:g\bar u u:>^2$ \cite{SVZ79} and the
estimation for $m_0^2$  \cite{BI82}.

The expressions for $Q_{ax}(x)$ and $D_{ax}(x)$ may be considered as
generation functions to obtain the condensates of higher dimensions
in the SI approximation. From a technical point of view this
procedure is more convenient than the direct calculations of them. In
Appendix A we present some new relations for quark VEVs of dimension
seven and gluon VEVs of dimension eight in the SI approximation.

\section{Nonlocal condensates within the single instanton approximation.}
The aim of this section is to study the form of the distributions
over virtuality of quarks and gluons in the SI approximation.  To
understand the main asymptotical behavior of correlators at short and
long distances it is enough to inspect the expressions \re{E18} and
\re{Mg} dropping the Schwinger $\hat E$  factor. We will also
consider numerical effects connected with the neglect of this factor.

To this goal, let us first calculate the correlators using the
regular gauge and neglecting  the $\hat E$ factor.  The corresponding
expressions are given by \re{E24} and \re{Dgax} with the changes
$\cos(...) \to 1$ and $\sin(...) \to 0$ in the integrands and are
reduced to
\ba
\ds Q^{reg}(x^2) & = &\ds \frac{2}{y^2}
\left( 1-\frac{1}{\sqrt{1+y^2}} \right),
\label{Q_reg} \\
D^{reg}(x^2) & = & \ds \frac{3}{4y^2(1+y^2)}
\left(\frac{1+2y^2}{y\sqrt{1+y^2} } \ln|\sqrt{1+y^2}+y| -1 \right),
\label{Q_reg_g}
\ea
where the dimensionless parameter $\ds y = \frac{x}{2\rho}$ is
introduced.

From \re{Q_reg} and \re{Q_reg_g} we easily find for the average
virtualities
$$
\lambda^2_{q,reg} = \frac{3}{2}\frac{1}{\rho^2},\ \ \  \ \ \
\lambda^2_{g,reg} = \frac{16}{5}\frac{1}{\rho^2},
$$
which are about $30\%$ less than the corresponding gauge - invariant
``physical" values in (\ref{E25}).  The same quantities (without
$\hat E$ factor) calculated in the singular gauge  look like
\be
\lambda^2_{q,sing} = \frac{9}{2}\frac{1}{\rho^2},\ \ \  \ \ \
\lambda^2_{g,sing} = \frac{96}{5}\frac{1}{\rho^2}. \la{sing}
\ee
Thus, we see that the gauge dependence is very strong and the results
derived without  $\hat E$- factors may be far from being correct
numerically\footnote{Note, that an estimate for $\lambda^2_{q}$
calculated by non-gauge invariant manner
(which is close to $\lambda^2_{q,sing}$ in (\ref{sing})) is presented
in \cite{KO94}.}.

Now, let us consider the correlation functions $f(\nu)$ in the
regular gauge.  To this end, we make the inverse Laplace transform of
the correlators \re{Q_reg} and \re{Q_reg_g} and obtain
\ba
\label{E26}
f_{q}^{reg}(\nu)&=&2\rho^2 \cdot \mbox{erfc}(\rho\sqrt{\nu}),\\
\label{laplreg}
f_{g}^{reg}(\nu)&=&\frac{3}{2}\rho^2 \cdot \left( \frac{\rho^2 \nu}{2}\right)
\exp\left(-{\rho^2 \nu \over 2}\right)  K_0\left({\rho^2 \nu \over 2}\right),
\ea
where erfc$(t) = 1-$erf$(t)$ is the error function and $K_{0}(t)$ is
the Mac-Donald function.  Then, it is easy to obtain large $\nu$
asymptotics of these functions
\ba
f_{q}^{reg}(\nu)&=&2\rho^2
\frac{e^{-\rho^{2}\nu}}{\sqrt {\rho^2\nu\pi}}
\left(1+O\left(\frac{1}{\nu}\right)\right)
\label{f_reg_q},  \\
f_{g}^{reg}(\nu)&=&\frac{3}{4}\rho^2
e^{-\rho^{2}\nu}\sqrt{\rho^2\nu\pi}\left(1+O\left(\frac{1}{\nu}\right)\right),
\label{f_reg_g} \ea
which reflect the behavior of the corresponding correlators in the
region of small $x$.  The same exponential asymptotics have physical
correlation functions $f_{q(g)}(\nu)$ resulting from the
gauge-invariant correlators \re{E24} and \re{Dgax}. Thus, we can
conclude that the model of nonlocal condensates in the SI
approximation can reproduce the main exponential asymptotical
behavior $\sim \exp(- \sigma \cdot \nu)$ of the physical correlation
functions at large virtualities (short distances), and the
phenomenological parameter $\sigma$ in Exp. (\ref{E22}) may be
identified as $\sigma \simeq \rho_c$.

As to the description of the small virtuality (long distance) region,
this approximation fails since in that regime $f(\nu\to 0)$ decays
too slowly in contradiction with the physically argued ``color
screening" exponential asymptotics given in \re{fsmall}.  In other
words, the correlators in (\ref{Q_reg}) and (\ref{Q_reg_g}) decrease
 too slowly at large $x$. These conclusions remain valid for the
physical case of gauge-invariant correlators \re{E24} and \re{Dgax},
that is easily seen from the behavior of corresponding numerators of
the integrands at large $x$.  As it is explained in \cite{DEMM296},
the SI approximation considered in the present paper does not
correspond to the real physical vacuum picture.  We should take into
account the important long-wave background fields too.  These fields
modify the long-distance behavior of the correlators and lead to
appearance of the ``second scale" parameters $\Lambda_q$ and
$\Lambda_g = 1/L_g$ in quark and gluon distributions, respectively
(see Exp.(\ref{fsmall}) and discussion there).  This effect allows us
to reproduce the long- and short-distance behavior of the physical
correlators \re{E22} in a complete form.  It is also shown that the
effect of long-wave vacuum fluctuations is not very essential for the
values of $\lambda_{q(g)}$ related to short distances.

\section{Conclusion}
The instanton model provides a way for constructing of the nonlocal
vacuum condensates. We have obtained the expressions for the nonlocal
gluon $<:Tr G^{\mu\nu}(x)\hat{E}(x,y)G^{\rho\sigma}(y)\hat{E}(y,x):>$
and quark \\ $<:~\bar {q}(0)\hat{E}(0,x)q(x):>$  condensates within
the single instanton approximation. The average virtualities of
quarks $\lambda_q^2$ and gluons $\lambda_g^2$ in the QCD vacuum are
derived.  The results are $\ds \lambda_q^2 = 2\frac{1}{\rho_c^2}$ for
vacuum quarks, and $\ds \lambda_g^2 = \frac{24}{5}\frac{1}{\rho_c^2}$
for vacuum gluons.  The value of $\lambda_q^2$ estimated in the QCD
SR analysis \cite{BI82} is reproduced at $\rho_c \approx 2\ {\rm
GeV}^{-1}$. This number is close to the estimate from  the
phenomenology of the QCD vacuum in the instanton liquid model
\cite{Shuryak96,PDW96}.  The model provides parameterless  predictions for
the ratio $\lambda_g^2/\lambda_q^2 = 12/5$ and the relation
(\ref{dim6}) for the vacuum averages of dimension six.

The calculations have been performed in a gauge - invariant manner by
using the expressions for the instanton field and quark zero mode in
the axial gauge~\cite{MHax}. It is shown that the usage of the
singular gauge (in neglecting the Schwinger gauge factor
$\hat{E}(0,x)$) in the calculations of non - gauge - invariant
quantities leads to a strong numerical deviation from correct values.

The behavior of the correlation functions demonstrates that in the
single instanton approximation the model of nonlocal condensates can
well reproduce the asymptotic behavior of the functions \re{E22}
at large virtualities (short distances).
The results for nonlocal quark \re{E24} and gluon \re{Dgax}
condensates are supported by the independent test of their local
characteristics $\lambda_q^2$ and $\lambda_g^2$. The latter may be
obtained from standard VEVs calculated within the instanton model in
\cite{SVZ79}, \cite{Sh82} and  \cite{PolW96}.

Nevertheless, the approximation used fails in the description of the
small virtuality (long distance) regime. The reason is the neglect of
long - wave vacuum fluctuations in the single instanton
approximation.  In the forthcoming paper \cite{DEMM296} we will prove
that inclusion of the effects of these fluctuations cures this
disease.

\vspace{1cm}

\centerline{\bf Acknowledgments}
\vspace{2mm}
The authors are grateful to Drs. H. Forkel, M. Hutter,
N.I. Kochelev, A.E. Maximov,
M.V. Polyakov, and R. Ruskov for fruitful discussions of the
results, and Prof. A.V. Radyushkin for clarifying remarks.
One of us (A.E.D.) thanks Wuppertal University and Prof. P.
Kroll for warm hospitality.  This investigation has been supported in
part by the Russian Foundation for Fundamental Research (RFFR)
96-02-17631 (S.V.M.), 96-02-18096 (A.E.D., S.V.E) and 96-02-18097
(A.E.D.) and INTAS 93-283-ext (S.V.E.).

\begin{appendix}
\appendix
\section{Appendix}

Here, we present the relations between the derivatives
$\ds \frac{d^2 Q(x^2)}{2! d^2 x^2}$ and $\ds \frac{d^2 D(x^2)}{2! d^2 x^2}$
calculated in the SI approximation
and the same quantities expressed via the quark and gluon VEVs obtained in
\cite{Groz} (below we put the current quark mass $m_q =0$)
\ba
\frac{d^2 Q_{ax}(x^2)}{2! d^2 x^2} & \equiv & \frac{7}{120}\frac{1}{\rho_c^4}
 =  \frac{3 Q^7_1 - \frac{3}{2} Q^7_2 - 3 Q^7_3 + Q^7_4 }{4! 24 <\bar q q>}
~~~~~~~~~~(\mbox{SI approximation}),
\la{dQ4}\\
\frac{d^2 D_{ax}(x^2)}{2! d^2 x^2} & \equiv  & \frac{5}{21}\frac{1}{\rho_c^4}
= \frac{\frac{7}{2} G^8_{1-2} + 23 G^8_{3-4} + 30 G^8_5 + 8 G^8_6 - 3 G^8_7 }
{4! 6 G^4}~~(\mbox{SI approximation}),
\la{dD4}\ea
where the quark condensate basis was chosen in the form
\ba
Q^7_1 & = & <\bar q G_{\mu\nu}G_{\mu\nu} q>,\ \ \ \
Q^7_2 = i <\bar q G_{\mu\nu}\tilde G_{\mu\nu} \gamma_5 q>,
\nn\\
Q^7_3 & = & <\bar q G_{\mu\lambda}G_{\lambda\nu} \sigma_{\mu\nu} q>,\ \ \ \
Q^7_4 = i <\bar q D_{\mu}J_\nu \sigma_{\mu\nu} q>,
\nn\ea
and the gluon condensate basis was chosen as
\ba
G^4 & = & <Tr G_{\mu\nu}G_{\mu\nu}>,\nn\\
G^8_1 & = &  <Tr G_{\mu\nu}G_{\mu\nu} G_{\alpha\beta}G_{\alpha\beta}>,\ \ \
G^8_2 = <Tr G_{\mu\nu} G_{\alpha\beta}G_{\mu\nu} G_{\alpha\beta}>,\ \ \
\nn\\
G^8_3 & = &  <Tr G_{\mu\alpha} G_{\alpha\nu}G_{\nu\beta} G_{\beta\mu}>,\ \ \
G^8_4 = <Tr G_{\mu\alpha} G_{\alpha\nu}G_{\mu\beta} G_{\beta\nu}>,
\nn \\
G^8_5  & = & i <Tr J_\mu G_{\mu\nu} J_\nu >,\ \ \
G^8_6 = i <Tr J_\lambda [ D_\lambda G_{\mu\nu}, G_{\mu\nu}] >,\ \ \
G^8_7 = <Tr J_\mu D^2 J_\nu >,
\nn\ea
and  the notation $G^8_{i-j} = G^8_{i} - G^8_{j}$ is used.

\end{appendix}


\begin{thebibliography}{99}
\bibitem{SVZ79} M.A. Shifman, A.I. Vainshtein. and V.I. Zakharov     
{\it Nucl.Phys} {\bf B147} (1979) 385, 448.

\bi{MihRad92} S.V. Mikhailov and A.V. Radyushkin {\it Sov. J. Nucl. Phys.}
        {\bf 49} (1989) 494; {\it Phys. Rev.} {\bf D45} (1992) 1754.
\bi{BI82}  V.~M. Belyaev and B.L.Ioffe  {\it Sov. Phys. JETP} {\bf 56} 
(1982) 493 ({\it Zh. Eksp. Teor. Fiz.} {\bf 83} (1982) 876;\\
  A.~A. Ovchinnikov, A.~A. Pivovarov, {\sl Yad. Fiz.} {\bf 48} (1988) 1135.

\bi{KSch87} M.Kramer, G. Schierholtz
{\it Phys. Lett.} {\bf B194} (1987) 283.                                

\bibitem{Sh82} E.V. Shuryak {\it Nucl. Phys.} {\bf B203} (1982) 93; 116.
\bi{PolW96} M.V. Polyakov, C. Weiss {\it Phys. Lett.} {\bf B387} (1996)
841.                                                                     

\bi{DiGicor96} A. Di Giacomo, E. Meggiolaro, H. Panagopoulos          
{\it Pisa University preprint IFUP - TH 12/96, UCY - PHY -
  96/5}, Pisa, 1996; hep-lat/9603017; {\it QCD96 Conference
  talk, Montpellier}, hep-lat/9608008;\\
 A. Di Giacomo, H. Panagopoulos {\it Phys. Lett.} {\bf B285} (1992) 133.


\bi{Mih93} S.V. Mikhailov {\it Phys. At. Nucl.} {\bf 56} (1993) 650.      

\bibitem{R1} A.M. Polyakov {\it Phys. Lett.} {\bf B59} (1975) 82;\\
~A.A.~Belavin, A.~M. Polyakov, A.A.~Schwartz and
Yu.S. Tyupkin {\it Phys. Lett.} {\bf B59} (1975) 85.                    

\bibitem{R4} G. 't Hooft  {\it Phys. Rev. } {\bf D14} (1976) 3432.         

\bibitem{Sh88} E.V. Shuryak {\it Nucl. Phys.} {\bf B302} (1988) 559.  

\bibitem{DP84} D.I. Dyakonov, V.Yu. Petrov, {\it Nucl. Phys.} {\bf
    B245} ( 1984) 259; {\it Nucl. Phys.} {\bf B272} ( 1986) 457.           
\bi{Shuryak96} T. Schafer, E.V. Shuryak, (1996), review,
hep-ph/9610451 and references therein.                               

\bibitem{Sh83} E.V. Shuryak {\it Nucl. Phys.} {\bf B214} (1983) 237.       

\bibitem{DKsr90} A.E. Dorokhov, N.I. Kochelev {\it Z. Phys.}
{\bf C46} (1990) 281.                                                      

\bi{DEMM296}  A.E. Dorokhov, S.V. Esaibegyan, A.E. Maximov,
S.V. Mikhailov, {\it in preparation}.                                

\bi{DoSi88} H.G. Dosch, Yu.A. Simonov {\it Phys. Lett.} {\bf B205}
(1988) 339.                                                                
\bi{NRad83} S.N. Nikolaev, A.V. Radyushkin
{\it Nucl. Phys.} {\bf B213} (1983) 285.
\bi{Groz} A.G. Grozin {\it Int.J. Mod. Phys.} {\bf A 10} (1995) 3497.         

\bibitem{Rad91} A.~V.~Radyushkin {\it Phys.Lett.} {\bf B271} (1991) 218.   


\bibitem{BM95} A.V. Radyushkin {\it CEBAF preprint TH - 94 - 13} 1994,
Newport News, hep-ph/9406237;\\
A.P. Bakulev, S.V. Mikhailov {\it Z. Phys.}{\bf C68}(1995) 51.

\bibitem{BM96} A.~P.~Bakulev and S.~V.~Mikhailov
	{\it Mod. Phys. Lett.} {\bf A11} (1996) 1611. 

\bibitem{MHax} M. Hutter {\it Munchen preprint LMU-95-03}
(1995), hep-ph/9502361.                                                       

\bi{BP82} V.N. Baier, Yu.F. Pinelis
{\it Phys. Lett.} {\bf B116} (1982) 179.

\bibitem{PDW96} D.I. Diakonov, M.V. Polyakov, C. Weiss
{\it Nucl. Phys.} {\bf B461} (1996) 539.

\bi{KO94} N. Kodama, M. Oka {\it Phys. Lett.} {\bf B340} (1994) 221.

\end{thebibliography}
\end{document}